\documentclass[letter]{aa} 

\usepackage{graphicx}
\usepackage{natbib} 
\bibpunct{(}{)}{;}{a}{}{,} 

\usepackage[varg]{txfonts}

\newcommand{\kms}{km\,s$^{-1}$}
\newcommand{\um}{$\mu$m}

\newcommand{\lsol}{L$_{\odot}$}
\newcommand{\msol}{M$_{\odot}$}
\newcommand{\msolyr}{M$_{\odot}$\,yr$^{-1}$}

\begin{document}

   \title{First image of the L1157 molecular jet \\
by the CALYPSO IRAM-PdBI survey}

   \author{L. Podio
          \inst{1}
          \and
         C. Codella
          \inst{1}
          \and
          F. Gueth
          \inst{2}
          \and
          S. Cabrit
          \inst{3,4,5}
          \and
          A. Maury
          \inst{6}
          \and
         B. Tabone
          \inst{3}
          \and
         C. Lef\`evre
          \inst{2}
          \and
         S. Anderl
          \inst{4,5}
          \and
         P. Andr\'e
          \inst{6}
          \and
         A. Belloche
          \inst{7}
          \and
         S. Bontemps
          \inst{8}
         \and
         P. Hennebelle
          \inst{6}
          \and
         B. Lefloch
          \inst{4,5}
          \and
         S. Maret
          \inst{4,5}
          \and
         L. Testi
          \inst{9,1}
          }

   \institute{
     INAF - Osservatorio Astrofisico di Arcetri, Largo E. Fermi 5, 50125 Firenze, Italy \\ 
     \email{lpodio@arcetri.astro.it}
   \and 
   IRAM, 300 rue de la Piscine, 38406 Saint Martin d’H\`eres, France
   \and
   LERMA, Observatoire de Paris, PSL Research University, CNRS, UMR 8112, F-75014, Paris France
   \and
   Univ. Grenoble Alpes, IPAG, F-38000 Grenoble, France
   \and
   CNRS, IPAG, F-38000 Grenoble, France
   \and
   Laboratoire AIM-Paris-Saclay, CEA/DSM/Irfu – CNRS – Universit\'e Paris Diderot, CE-Saclay, 91191 Gif-sur-Yvette, France
   \and
   Max-Planck-Institut f\"ur Radioastronomie, Auf dem H\"ugel 69, 53121 Bonn, Germany
   \and
   OASU/LAB-UMR5804, CNRS, Universit\'e Bordeaux 1, 33270 Floirac, France
   \and
   ESO, Karl-Schwarzschild-Strasse 2 D-85748 Garching bei M\"unchen, Germany
             }

   \date{Received ; accepted }

 
  \abstract
{Fast jets are thought to be a crucial ingredient of star formation because they might extract angular momentum from the disk and
thus allow mass accretion onto the star. However, it is unclear whether jets are ubiquitous, and likewise, their contribution to mass and angular momentum extraction during protostar formation
remains an open question.}
{Our aim is to investigate the ejection process in the low-mass Class 0 protostar L1157.
This source is associated with a spectacular bipolar outflow, and the recent detection of high-velocity SiO suggests the occurrence of a jet.
}
{Observations of CO~$2-1$ and SiO~$5-4$ at $\sim0\farcs8$ resolution were obtained with the IRAM Plateau de Bure Interferometer (PdBI) as part of the CALYPSO large program. The jet and outflow structure were fit with a precession model. We derived the column density of CO and SiO, as well as the jet mass-loss rate and mechanical luminosity.}
{High-velocity CO and SiO emission resolve for the first time the first 200~au of the outflow-driving molecular jet. The jet is strongly asymmetric, with the blue lobe $\sim0.65$ times slower than the red lobe.
This suggests that the large-scale asymmetry of the outflow is directly linked to the  jet velocity and that the asymmetry in the launching mechanism has been at work for the past $1800$~yr. Velocity asymmetries are common in T Tauri stars, which suggests
that the jet formation mechanism from Class 0 to Class II stages
might be similar.
Our model simultaneously fits the properties of the inner jet and of the clumpy 0.2~pc scale outflow by assuming that the jet precesses counter-clockwise on a cone inclined by $73\degr$ to the line of sight with an opening angle of $8\degr$ on a period of $\sim1640$~yr. 
The estimated jet mass flux and mechanical luminosity  are $\dot{M}_{\rm jet}\sim 7.7\times10^{-7}$~\msolyr\, and $L_{\rm jet}\sim0.9$~\lsol, indicating that the jet could extract at least 25\% of the gravitational energy released by the forming star.} 
{}


\titlerunning{First image of the L1157 molecular jet by CALYPSO}

               \maketitle
%

\section{Introduction}
\label{sect:intro}

Young Class~0 protostars ($t<10^5$ yr)
are representative of the main accretion phase, during which most of the final stellar mass is accreted from the dense parental envelope onto the central protostellar embryo \citep{andre00}. During this phase, the gas in the protostellar envelope must reduce its specific angular momentum by 5 orders of magnitude to allow the star to form without reaching its break-up speed. 
Early theoretical studies \citep[e.g., ][]{pudritz86} suggested that this angular momentum problem might be solved thanks to the ejection of magneto-centrifugal jets, which transport away angular momentum.

Fast and collimated molecular jets as well as slower wide-angle outflows, which are thought to be produced by the ambient gas entrained and accelerated by the ejected material, are frequently observed at the earliest protostellar stage \citep[e.g., ][]{arce07}. 
However, several aspects are still unclear, for instance, the
amount of accretion energy and angular momentum that the jet extracts from the system and the fraction of time for which they are active.
Most importantly, it is not clear whether jets are truly universal or only present in the most powerful outflow sources.

\begin{figure*}
\centering
  \includegraphics[width=4.cm]{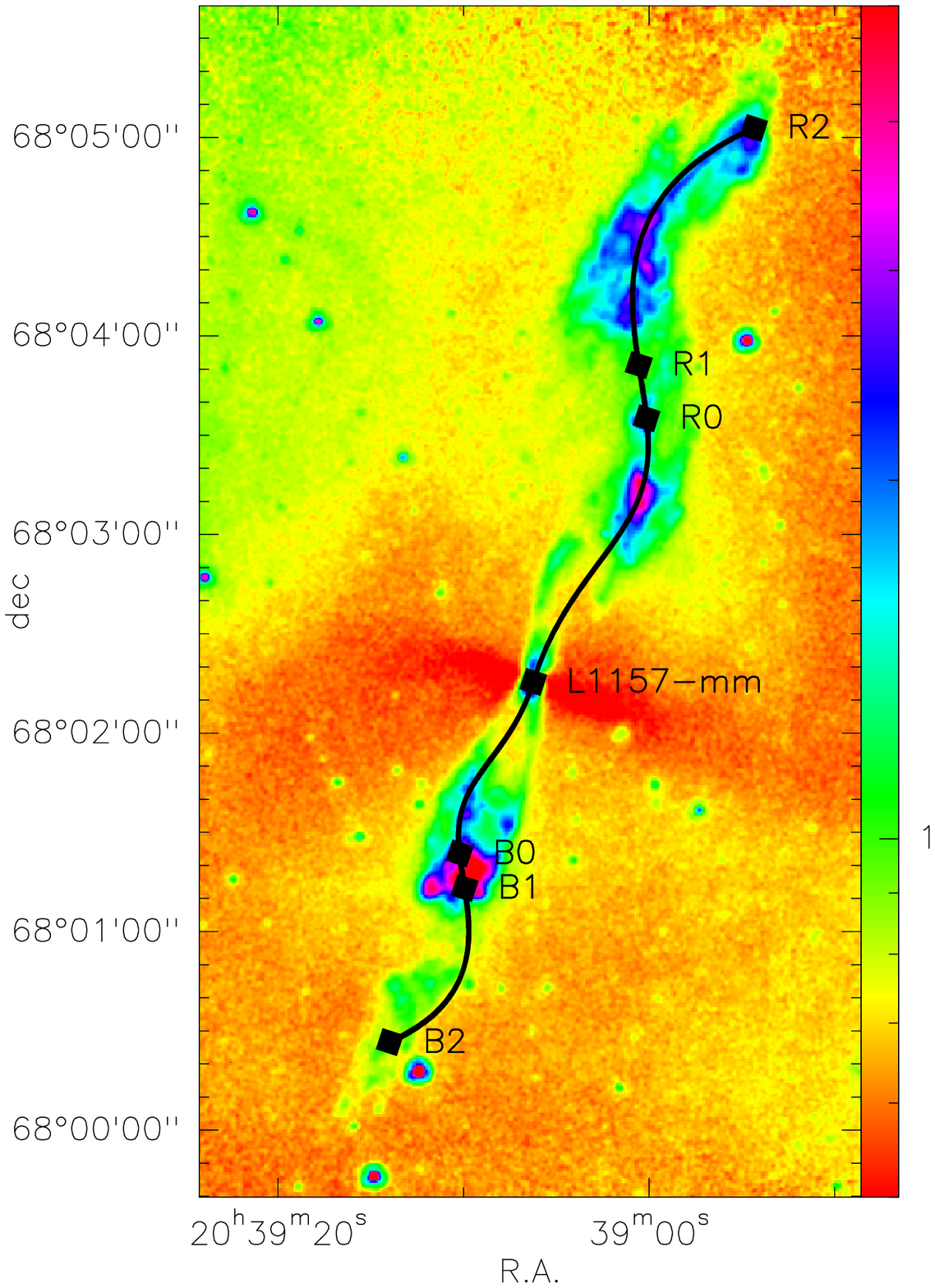}
  \includegraphics[width=13.cm]{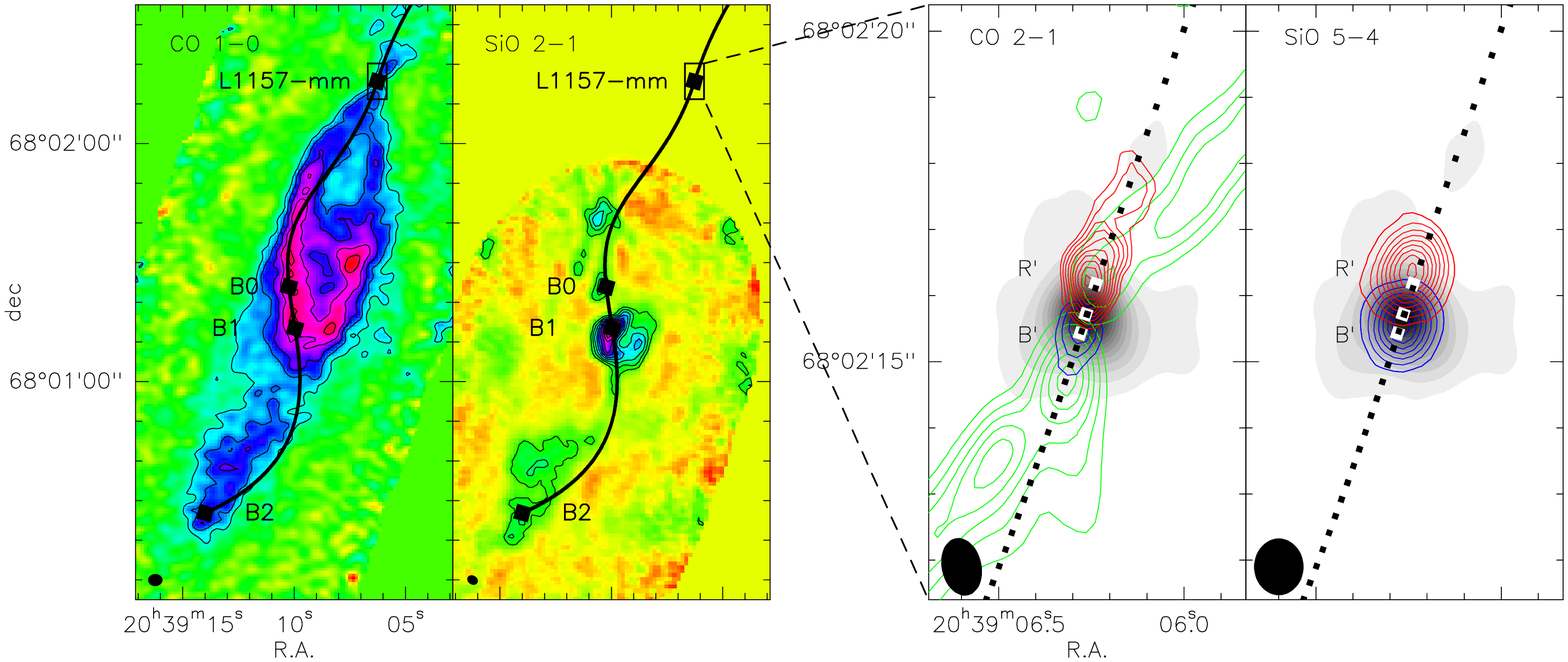}
  \caption{
{\it Left panel:} Spitzer/IRAC 8 \um\, observations of the L1157 outflow \citep{looney07,takami11}. The resolution is $\sim 1\farcs98$ and the intensity color scale is logarithmic from 0.05 to 5 mJy/sr. The precession model (black line) is overplotted.
{\it Middle panels:} the precession model (black line) is overplotted on integrated CO $1-0$ and SiO $2-1$ maps of the L1157 outflow by \citet{gueth96,gueth98} (IRAM-30m + PdBI data, $|V-V_{\rm sys}|= 0-11$ \kms\, for CO and $0-18$ \kms\, for SiO, HPBW: $3\farcs6 \times 3\farcs0$, PA=90$\degr$ for CO, $2\farcs8 \times 2\farcs2$, PA=56$\degr$ for SiO). The first contour is at 5$\sigma$  with steps of 3$\sigma$ (1$\sigma = 0.5$ Jy \kms/beam for CO,  and 0.2 Jy \kms/beam for SiO). The black squares indicate the driving source, L1157-mm, and the blueshifted knots.
{\it Right panels:} zoom-in on the inner jet revealed by CO $2-1$ and SiO $5-4$  maps obtained with IRAM-PdBI as part of CALYPSO. The emission is integrated over high blue- and redshifted velocities (in $|V-V_{\rm sys}|$) (blue/red contours) (HV$_{\rm b} = 22-57$ \kms, HV$_{\rm r}=36-63$ \kms). CO $2-1$ also shows emission at low velocities (green contours) (LV$_{\rm b}< 22$ \kms, LV$_{\rm r}< 36$ \kms) . The gray-scale traces the continuum emission at 1.3mm. The white squares indicate the continuum peak and the inner HV knots B' and R'. The ellipse in the bottom left corner shows the beam HPBW ($0\farcs88 \times 0\farcs60$, PA=12$\degr$ for CO, $0\farcs85 \times 0\farcs75$, PA=0$\degr$ for SiO). The first contour is at 5$\sigma$  with steps of 3$\sigma$ (1$\sigma \sim$ 1.3 mJy/beam for continuum,  and $\sim$ 50, 70, 180 mJy \kms/beam for SiO, CO HV, and CO LV, respectively). The precession is plotted in steps of $0.0025 \times P$, i.e., $\sim 4.1$ yr, where $P=1640$ yr is the precession period (black squares).}
  \label{fig:l1157_prec_pdbi}
\end{figure*}

A particularly interesting target in which to investigate the role of jets at the early protostellar stage is the low-mass Class 0 source L1157-mm, located at $d$ $\simeq$ 250 pc \citep{looney07}.
The protostar is associated with a spectacular bipolar outflow that has been the subject of a large number of observational campaigns. These studies have shown a number of indirect pieces
of evidence of a jet driven by the L1157-mm: (i) a chain of curved bow shocks in NH$_3$, SiO, H$_2$, indicating the action of a precessing time-variable jet \citep{tafalla95,zhang95,gueth98,takami11}; (ii) elongated CO cavities trailing behind these bright shocks, suggestive of jet-driven entrainment in the bow wings \citep{gueth96}; (iii) and a compact dissociative shock in [$\ion{o}{i}$] 63 $\mu$m behind the tip of the brightest bow shock B1, consistent with a jet Mach disk with momentum flux sufficient to drive the large-scale CO outflow \citep{benedettini12}. 
Different scenarios have been proposed to explain how such a jet can produce the morphology and kinematics of the spectacular parsec-scale outflow. 
\citet{gueth96} and \citet{bachiller01}  suggested that the observed outflow cavities are opened by a jet precessing on the surface of a cone with a period of about $4000-5000$ yr. \citet{kwon15} proposed instead that the curved CO structures are not tracing swept-up cavities but a pair of precessing, misaligned jets.
However, so far no jet was imaged close to the source in mm/sub-mm interferometric maps.
Only very recently, \citet{tafalla15} reported the detection of SiO high-velocity emission by using single-dish IRAM-30m observations, supporting the occurrence of a jet in the inner 11$\arcsec$ around L1157-mm.

In this Letter we report the first images of the L1157 jet, using SiO and CO emission, in the framework of the IRAM PdBI large program continuum and lines in young protostellar objects (CALYPSO)\footnote{http://irfu.cea.fr/Projects/Calypso} . The presented observations allow us to derive the jet properties and its interplay with the large-scale outflow.

\section{Observations and data reduction}
\label{sect:obs}

The central $20\arcsec$ region toward the L1157-mm protostar  was observed with the IRAM PdBI in February, March, and November 2011 using the A and C configurations.
The shortest baseline is $\sim 20$ m and the longest is $\sim 760$ m, allowing us to recover emission at scales from $\sim 8 \arcsec$ to $\sim 0\farcs8$ at 1.3 mm.
The SiO $5-4$ and CO $2-1$ lines at 217104.98 and 230538.00 MHz were observed using the WideX backend, which covers a 4 GHz spectral window with a spectral resolution of $\sim$2 MHz, which is $\sim 3.2-3.4$ \kms\, at the considered wavelengths.
The calibration was carried out with GILDAS-CLIC\footnote{http://www.iram.fr/IRAMFR/GILDAS} following standard procedures.
The phase rms was $\le 50\degr$ and $\le 80\degr$ for the A and C tracks, respectively, the pwv was $0.5-1$ mm (A) and $\sim1-2$ mm (C), and the system temperature was $\sim 100-160$ K (A) and $150-250$ K (C). 
The uncertainty on the absolute flux calibration is $\le 15$\% and the typical rms noise per spectral resolution channel $\sim 3-9$ mJy/beam.
Images were produced using robust weighting, resulting in a clean beam of $\sim 0\farcs85 \times 0\farcs75$ (PA=0$\degr$) for SiO $5-4$ and $\sim 0\farcs88 \times 0\farcs60$ (PA=12$\degr$) for CO $2-1$.

\section{Asymmetric molecular jet}
\label{sect:jet}

Figure \ref{fig:l1157_prec_pdbi} shows in the left panel a H$_2$ Spitzer map of the L1157 outflow reported by \citet{looney07}, in the middle panel IRAM-PdBI maps of the blueshifted outflow lobe in CO $1-0$ and SiO $2-1$ at $\sim 3\arcsec$ resolution  \citep{gueth96,gueth98}, and in the right panel a zoom-in on the central 1000 au around the protostar in CO $2-1$ and SiO $5-4$ at much higher resolution ($\sim 0\farcs8$) obtained from our IRAM-PdBI CALYPSO observations.
Both CO $2-1$ and SiO $5-4$ show compact emission at high blue- and redshifted velocities, probing a molecular jet.
The high-velocity range (HV) is defined based on SiO $5-4$, which solely shows emission at high velocities, that is, at $22<|V-V_{\rm sys}|< 57$ \kms\, in the blue lobe and at $36<|V-V_{\rm sys}|< 63$ \kms\, in the red lobe (the systemic velocity in the LSR is $V_{\rm sys} = +2.7$ \kms, \citealt{bachiller97}).
CO $2-1$ also shows low-velocity emission (LV) with an X-shaped pattern (with one arm of the X much brighter than the other), coinciding with the walls of the CO cavity shown on larger scale by \citet{gueth96}. The HV compact emission is oriented roughly along the middle of this X, which suggests that the LV CO traces ambient material that is swept-up by a single precessing jet, as proposed by \citet{gueth96}, and not two misaligned jets (see \citealt{kwon15}). 



The HV CO $2-1$ and SiO $5-4$ emission reveals for the first time the base ($\sim 200$ au) of the bipolar high-velocity jet driving the large-scale outflow, in agreement with what was recently suggested by \citet{tafalla15} based on spatially unresolved IRAM-30m observations of the SiO $5-4$ and $6-5$ lines.
The molecular jet consists of a blue- and a redshifted knot (named B' and R') located along PA$_{\rm jet}\sim-17\degr$ with an emission peak at $\sim 0\farcs22$ and $\sim 0\farcs61$ distance from L1157-mm, respectively.
The position velocity diagram of the CO $2-1$ and SiO $5-4$ emission extracted along the jet PA is shown in Fig.~\ref{fig:l1157_pv}.
The jet width is estimated by a Gaussian fit of the CO intensity profile perpendicular to the jet axis (the jet is not resolved transversally in SiO because of the larger beam size).
CO has a FWHM at the emission peak of $\sim 0\farcs77$ in the redshifted knot and $\sim 0\farcs9$ in the blueshifted knot. This corresponds to intrinsic widths of $\sim 0\farcs25$ (i.e., $\sim 60$ au), and $\sim 0\farcs5$ (i.e. $\sim 125$ au), respectively, after correcting for beam smearing (the effective resolution across the jet is $HPBW_{\rm tr} \sim 0\farcs73$, equivalent to the geometric mean of the beam major and minor axes). 
This indicates a strong jet asymmetry where the blueshifted knot is less collimated and less extended than the redshifted knot.

\begin{figure}
\centering
  \includegraphics[width=5.6cm]{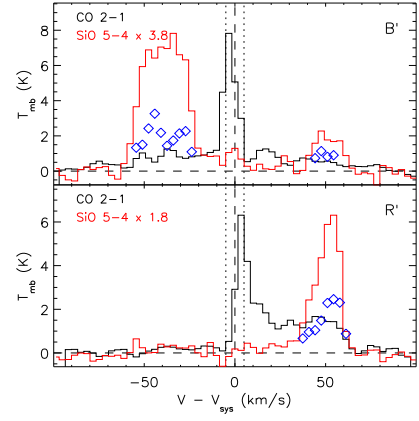}
  \caption{
Spectra of SiO $5-4$ (red) and CO $2-1$ (black), and their intensity ratios SiO $5-4$/CO $2-1$ (blue diamonds) at the position of the HV blue- ({\it top}) and redshifted ({\it bottom}) jet knots, B' and R'. The profiles are normalized to the CO peak intensity (normalization factor is labeled).
Horizonthal and vertical dashed lines mark the baseline and the systemic velocity, V$_{\rm sys} = +2.7$ \kms\ and the vertical dotted lines the $\pm 5$ \kms\, interval where extended CO emission is filtered out.}
  \label{fig:l1157_spec}
\end{figure}

Moreover, the spectra of SiO $5-4$ and CO $2-1$ in knots B' and R' (Fig. \ref{fig:l1157_spec}) indicate that the blueshifted knot is spectrally broader and has a lower peak radial velocity ($V_{\rm rad,b}\sim -35$ \kms) than the redshifted knot ($V_{\rm rad,r} \sim +55$ \kms). 
The observed jet asymmetry in velocity ($V_{\rm rad,b} / V_{\rm rad,r} \sim 0.65$) perfectly matches the asymmetry in length of the large-scale outflow (the precession model by \citealt{bachiller01} assumes exactly the same velocity ratio to reproduce the different length of the blue and red outflow lobes).
This indicates that the outflow asymmetry is directly linked to the jet velocity and is not due to a difference in the density of the interstellar medium (ISM).
The agreement with the different global lengths of the north and south lobes suggests that the asymmetry in the launching mechanism has been at work since the ejection of the terminal knots (B2 and R2).
Similar asymmetries are common in Class II sources \citep[e.g., ][]{hirth94,hirth97,melnikov09,podio11}.
In addition to the similarity in knot spacings and collimation between Class 0 and Class II jets \citep[e.g., ][]{cabrit07b,podio15}, this is another hint that the jet launching mechanism in protostars of $\sim 10^4$ yr might be similar to that for Class II sources ($\sim 10^6$ yr).

\section{Jet precession}
\label{sect:precession}

Previous studies \citep[e.g., ][]{gueth96,bachiller01} suggested that the spatial misalignment and slightly different radial velocities between the bright CO cavities in the blue and red lobes might be due to a jet slowly precessing on the surface of a cone inclined by an angle $i$ to the line of sight (l.o.s.) with an opening angle $\theta$ on a period $P$. Ejection events occurring while the jet precesses produce knots that then propagate with velocity $V_{\rm S}$. 
However, as the jet has never been mapped before, the model parameters ($i$, $\theta$, $V_{\rm S}$, precession sense, knot ages) were poorly constrained.
The properties of the inner jet knots revealed by our IRAM-PdBI observations set strong constraints on the precession model first presented by \citet{gueth96} as we describe below.\\
(i) The jet orientation near the protostar is PA$_{\rm jet} \sim-17\degr$. If the jet precesses, its orientation varies periodically around the PA of the precession cone (PA$_{\rm cone}\sim-22\degr$ to reproduce the large-scale outflow precession pattern, \citealt{gueth96}) between PA$_{\rm cone} - \theta$ and PA$_{\rm cone} + \theta$.
The PA of the most recent and closest knots indicates that the  precession axis currently is on the west side of the cone and that the opening angle should be $\theta \ge 5\degr$.\\
(ii) The jet radial velocity in the inner knot B', $\sim -35$ \kms, is higher than in knot B1 ($\sim -15 \pm 5$ \kms, \citealt{zhang95,gueth98,gomezruiz13}). Here, it is assumed that the HV SiO emission in B1 probes the jet impact region and not entrained material. This assumption is based on the fact that, while LV SiO has a bow-shock structure, the HV SiO ($\sim-8/-18$ \kms) is confined to a compact region \citep{gueth98}.
If the jet precesses and propagates with constant speed then the inclination with respect to l.o.s, $i_{\rm jet}$, varies periodically between $i - \theta$ and $i + \theta,$ as does the velocity projection along the l.o.s. ($V_{\rm rad}$). In this scenario,  $V_{\rm rad} (B') > V_{\rm rad} (B1)$ implies that the inclination at the time the knot B’ was ejected was smaller than for B1 ($V_{\rm rad} (B') / V_{\rm rad} (B1) = cos\,i_{\rm jet} (B') / cos\,i_{\rm jet} (B1)$),
meaning that the knots B' and B1 were  ejected when the precession axis was at the front and the back side of the precession cone, respectively.
This strongly constrains the inclination of the cone $i$ and the time of knot ejection. \\
(iii) The spectra of the inner knots, B' and R', indicate that the blueshifted jet travels at a speed $\sim 0.65$ times lower than the redshifted jet ($V_{\rm jet,b} = 0.65 V_{\rm jet,r}$).

To satisfy these constraints and reproduce the positions of knots B1 and B2 \citep{gueth96}, as well as the precession pattern in the northern lobe visible in the Spitzer map \citep{looney07}, we assume that the jet axis is precessing on a cone  inclined by $i = 73\degr$  with opening angle $\theta = 8\degr$, that the precession axis was at the front side of the cone at the time B2 was ejected ($i_{\rm jet} (B2) = 65\degr$) and precesses counter-clockwise (looking down from north), that the jet propagates with a constant speed $V_{\rm S}$, and that the ages of knots B2, B1, B0, and B' in the blue lobe (and the corresponding knots R2, R1, R0, and R' in the red lobe) relative to the period $P$ are $1.1 \times P$, $\sim 0.67 \times P$, $\sim 0.58 \times P$, and $\sim 0.002 \times P$ (which correspond to inclination angles of $65\degr$, $80.2\degr$, $81\degr$, and $66.4\degr$).
As shown in Fig. \ref{fig:l1157_prec_pdbi}, our precession model simultaneously fits the large-scale outflow (IRAM-PdBI CO and SiO maps by \citealt{gueth96,gueth98}  and H$_2$ Spitzer map by \citealt{looney07}) and the inner jet structure (i.e., the HV knots B' and R').

If the jet precession is due to disk precession induced by tidal interaction with a companion in a non-coplanar orbit, then the jet should precess in the opposite direction to the rotation of the disk (retrograde precession). This scenario is supported by C$^{18}$O observations by \citet{kwon15}, which indicate that the inner flattened envelope rotates clockwise.
Moreover, the modeling of PdBI visibilities from the 1.3 mm continuum emission suggests a $\pm70$ mJy secondary source at $<0\farcs3$ from L1157-mm (Maury et al. 2016 in prep.). 

\section{Jet properties}
\label{sect:mass_loss}

By correcting the observed radial velocities and tangential distances for the knot inclination angles, we derive the jet velocity, the distance and age of the knots, and the jet mass-loss rate and mechanical luminosity. 
The derived jet properties rely on the inclination of the precession cone inferred from the model ($i=73\degr$), which in turn depends on the ratio of the jet radial velocities in knots B' and B1. If the jet radial velocity at B1 is higher than what is inferred from SiO, for example, because of gas slow-down in the reverse shock, the inclination angle is smaller (e.g., $i\sim60\degr$ for $V_{\rm rad} (B1) = -23$ \kms), implying a jet velocity lower by $\sim30$\%. The small spatial overlap of the red and blue jet lobes close to the source and the flattened envelope in the Spitzer maps (Fig.~\ref{fig:l1157_prec_pdbi}) indicate, however, that the jet lies close to the plane of the sky, favoring large inclination angles. Therefore, we adopt $i=73\degr$. 
The deprojected velocities and distances of the inner knots B' and R' are $V_{\rm jet, b} \sim 87$ \kms, $V_{\rm jet, r}\sim 137$ \kms, and $d(\rm B')  \sim 60$ au, $d(\rm R')  \sim 166$ au.
If the B2 shock, located at a deprojected distance  $d(\rm B2)  \sim 130\arcsec$, propagates with a velocity $V_{\rm S} = V_{\rm jet, b} = 87$ \kms\, ,  then its dynamical timescale  is $\tau(\rm B2) = d(\rm B2) / V_{\rm jet, b} \sim 1800$ yr. 
This implies that the precession period is $P \sim 1640$ yr and the dynamical age of B1/R1, B0/R0, and B'/R' is $\sim 1100$, $\sim950$, and $\sim4$ yr.
The inferred knot ages are $\sim0.6$ times younger than previously estimated by \citet{gueth96}, mainly because of the assumed distance ($250$ pc instead of $440$ pc).

The beam-averaged column densities of CO and SiO in the jet, $N_{\rm CO}$ and $N_{\rm SiO}$, are derived from CO $2-1$ and SiO $5-4$ HV emission in knots B' and R' by assuming optically thin local thermodynamic equilibrium (LTE) emission at temperature $T_{\rm K} \sim T_{\rm ex} = 40 \pm 30$ K \citep{tafalla15}.
The ratio of SiO $5-4$ to CO $2-1$ in the HV range is $\sim2$ in the blue and red lobe, indicating similar SiO abundances (see Fig.~\ref{fig:l1157_spec}).
We obtain $N_{\rm CO} \sim 2.4^{+1.1}_{-0.6} \times 10^{16}$ cm$^{-2}$  in the blue and red knot and 
$N_{\rm SiO} \sim 8.2^{+1.8}_{-0.7} \times 10^{13}$ cm$^{-2}$ in B' and $\sim 7.5^{+1.8}_{-0.7} \times 10^{13}$ cm$^{-2}$ in R' (the uncertainty on $N_{\rm CO}$ and $N_{\rm SiO}$ is due to the assumed $T_{\rm ex}$).
The abundance of SiO is derived as $X_{\rm SiO} = X_{\rm CO} \times N_{\rm SiO}/N_{\rm CO}$, where $X_{\rm CO} \sim 10^{-4}$ is the CO abundance with respect to H$_2$. We find $X_{\rm SiO} \sim 3 \times 10^{-7}$ in the inner jet knots.

Then, the mass-loss rate of the molecular jet is calculated as $\dot{M}_{\rm jet} = m_{\rm H_2} \times (N_{\rm CO} / X_{\rm CO}) \times HPBW_{\rm tr} \times V_{\rm jet}$, where $V_{\rm jet}$ is the deprojected jet velocity,
$HPBW_{\rm tr}$ is the beam size across the jet ($\sim 0\farcs73$). 
We find $\dot{M}_{\rm jet,b} \sim 3.0^{+1.5}_{-0.7} \times 10^{-7}$ and $\dot{M}_{\rm jet,r} \sim 4.7^{+2.1}_{-1.2} \times 10^{-7}$  \msolyr. The inferred value for the blueshifted jet agrees with the estimate derived from the [\ion{o}{i}] 63\um\, luminosity in the B1 knot
($\sim 2 \times 10^{-7}$ \msolyr, \citealt{benedettini12}).

The determination of the jet speed and mass-loss rate allows us to estimate the jet mechanical luminosity, that is, $L_{\rm jet} = 1/2 \times \dot{M}_{\rm jet} \times V_{\rm jet}^2$.
We obtain $L_{\rm jet,b} = 0.2\pm0.1$ \lsol\, and $L_{\rm jet,r} = 0.7\pm0.2$ \lsol.
The total jet power is $\sim0.9$ \lsol, which is about one-fourth of the source bolometric luminosity ($\sim 3.6$ \lsol\, at $d=250$ pc, \citealt{green13}). 
This indicates that at least 25\% of the gravitational energy released by accretion onto the protostar could be extracted and converted into mechanical form in the jet.


Finally, an estimate of the outflow age, $\tau_{\rm out}$, can be derived by imposing that the momentum deposited by the jet in the blue lobe
$P_{\rm jet} = \dot{M}_{\rm jet} \times V_{\rm jet} \times \tau_{\rm out}$ is equal to the CO outflow momentum, that is, $P_{\rm out} =  M_{\rm CO} \times V_{\rm CO}$. 
The outflow momentum in the blue lobe, corrected for the assumed distance of 250 pc,  is $P_{\rm out} = 0.94$ \msol\, \kms \citep{bachiller01}.
As $\dot{M}_{\rm jet,b} = 3.0 \times 10^{-7}$ \msolyr\, and $V_{\rm jet,b} = 87$ \kms\, in the blueshifted jet, the outflow age is $\tau_{\rm out} \sim 3 \times 10^4$ yr, in agreement with the lifetime of Class 0 sources ($\sim 4-9 \times 10^4$ yr, see, e.g., \citealt{evans09,maury11}). 
The estimated outflow age is much older than the dynamical timescale of knot B2 inferred from our precession model ($\sim 1800$ yr).
The reason might be that the jet suddenly slows down when it
encounters much denser ambient gas in the farthest knots (B2 and R2), so that the CO outflow takes about $3 \times 10^4$ yr to reach its current total length. In this scenario only jet material ejected during the past $1800$ yr is visible in the CO cavity, while older ejecta have piled up at the tip of the CO outflow or against the cavity walls intercepted by the precessing jet (e.g., near B1). 
Alternatively, the discrepancy between $\tau_{\rm out}$ and $\tau(B2)$ might arise from older working surfaces that have moved far away from the source ($\sim 2.7$ pc, i.e., 36$'$, in $3 \times 10^4$ yr at 87 \kms) so that their CO counterparts have not been observed because of the limited mapping area, fainter CO emission and less efficient jet-entrainment at lower density, and increased confusion with static ambient cloud emission.
\section{Conclusions}
\label{sect:conclusions}

Our CALYPSO PdBI observations reveal for the first time the first 200 au of the fast and collimated ($\sim 60-125$ au) molecular jet driving the L1157 outflow. 
The observations indicate that the large-scale asymmetry in length of the molecular outflow is directly linked to an asymmetry in the jet velocity (the blue lobe is $\sim 0.65$ times slower then the red lobe). This also implies that the asymmetry in the launching mechanism has been at work for the past $1800$ yr.
The velocity asymmetry is similar to that frequently observed in Class II jets, suggesting that the same ejection process is at work throughout the star formation process from Class 0 to Class II sources.
Our model indicates that the clumpy large-scale outflow extending up to 0.2 pc might be driven by a jet precessing on the surface of a cone inclined by $73\degr$ to the l.o.s with an opening angle $\theta = 8\degr$ on a period $P\sim 1640$ yr. The precession is retrograde with respect to the envelope rotation, which favors tidal disk precession due to a non-coplanar companion.
The jet mass loss-rate is $\sim 7.7 \times 10^{-7}$ \msolyr\, and the mechanical luminosity is $\sim 0.9$ \lsol\, , indicating that at least 25\% of the gravitational energy released by the forming protostar might be dissipated by the jet.

\begin{acknowledgements}
L.P. has received funding from the European Union FP7, GA No. 267251.
The CALYPSO project has received further support from the European Union’s FP7 (ERC Advanced GA No. 291294 – `ORISTARS').
\end{acknowledgements}

\bibliographystyle{aa} 
\bibliography{../mybibtex} 

\clearpage

\begin{appendix}

\section{Position-velocity diagrams}

\begin{figure}
  \includegraphics[width=\columnwidth]{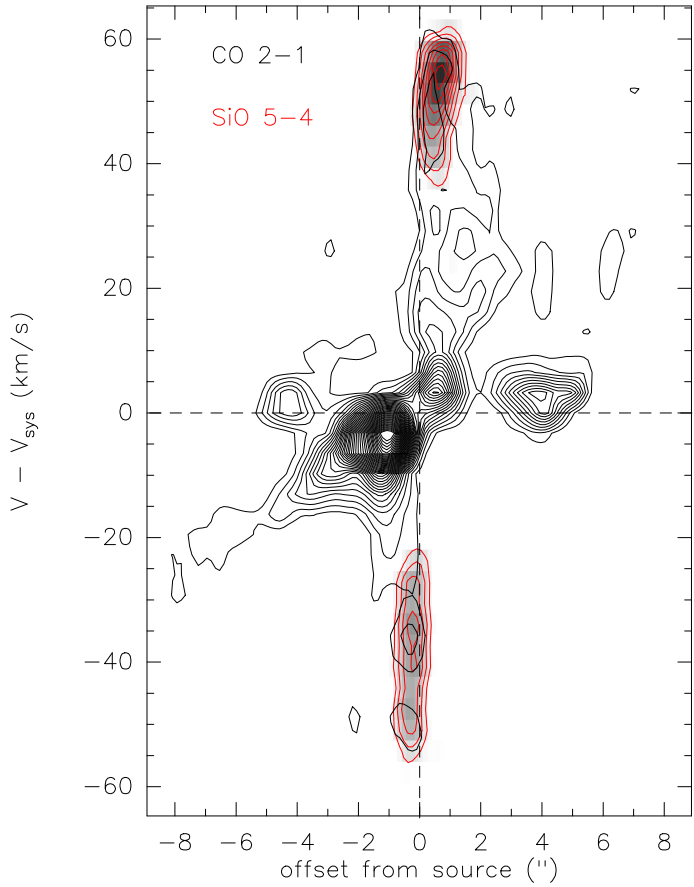}
  \caption{Position-velocity diagram of CO $2-1$ (black contours) and SiO $5-4$ (gray-scale and red contours) along the PA of the high-velocity molecular jet (PA$_{\rm jet}\sim-17\degr$).  Horizontal and vertical dashed lines mark the systemic velocity, V$_{\rm sys} = +2.7$ \kms, and the continuum peak.
The first contour is at 5$\sigma$ with steps of 3$\sigma$.}
  \label{fig:l1157_pv}
\end{figure}

The position velocity diagram of the CO $2-1$ and SiO $5-4$ emission extracted along the jet PA (PA$_{\rm jet}-17\degr$) is shown in Fig. \ref{fig:l1157_pv}.
The figure shows that the CO and SiO emission probe the same jet component at HV, that is, a compact and collimated jet close to the protostar.

\end{appendix}

\end{document}